\documentclass[11pt]{article} 

\usepackage[dvipsnames]{xcolor}
\usepackage[linktoc=page,bookmarks=false,colorlinks=false,
linkbordercolor=RoyalBlue,citebordercolor=ForestGreen,urlbordercolor=CornflowerBlue]{hyperref}

\usepackage[utf8]{inputenc}
\usepackage{graphicx}
\usepackage{amsmath}
\usepackage{mathabx}
\usepackage{authblk}
\usepackage{lineno}

\title{Search for secluded dark matter towards the Galactic Centre with the ANTARES neutrino telescope}

\author[1,2]{A.~Albert}
\author[3]{S.~Alves}
\author[4]{M.~Andr\'e}
\author[5]{M.~Anghinolfi}
\author[6]{G.~Anton}
\author[7]{M.~Ardid}
\author[7]{S.~Ardid}
\author[8]{J.-J.~Aubert}
\author[9]{J.~Aublin}
\author[9]{B.~Baret}
% ...
\author[ ]{

(The ANTARES Collaboration)}

\affil[1]{\scriptsize{Universit\'e de Strasbourg, CNRS,  IPHC UMR 7178, F-67000 Strasbourg, France}}
\affil[2]{\scriptsize Universit\'e de Haute Alsace, F-68100 Mulhouse, France}
\affil[3]{\scriptsize{IFIC - Instituto de F\'isica Corpuscular (CSIC - Universitat de Val\`encia) c/ Catedr\'atico Jos\'e Beltr\'an, 2 E-46980 Paterna, Valencia, Spain}}
\affil[4]{\scriptsize{Technical University of Catalonia, Laboratory of Applied Bioacoustics, Rambla Exposici\'o, 08800 Vilanova i la Geltr\'u, Barcelona, Spain}}
\affil[5]{\scriptsize{INFN - Sezione di Genova, Via Dodecaneso 33, 16146 Genova, Italy}}
\affil[6]{\scriptsize{Friedrich-Alexander-Universit\"at Erlangen-N\"urnberg, Erlangen Centre for Astroparticle Physics, Erwin-Rommel-Str. 1, 91058 Erlangen, Germany}}
\affil[7]{\scriptsize{Institut d'Investigaci\'o per a la Gesti\'o Integrada de les Zones Costaneres (IGIC) - Universitat Polit\`ecnica de Val\`encia. C/  Paranimf 1, 46730 Gandia, Spain}}
\affil[8]{\scriptsize{Aix Marseille Univ, CNRS/IN2P3, CPPM, Marseille, France}}
\affil[9]{\scriptsize{Universit\'e de Paris, CNRS, Astroparticule et Cosmologie, F-75013 Paris, France}}
% ...

\date{\today}

\begin{document}

\maketitle
\begin{abstract}
Searches for dark matter (DM) have not provided any solid evidence for the existence of weakly interacting massive particles in the GeV-TeV mass range.
Coincidentally, the scale of new physics is being pushed by collider searches well beyond the TeV domain. This situation strongly motivates the exploration of DM masses much larger than a TeV.
Secluded scenarios contain a natural way around the unitarity bound on the DM mass, via the early matter domination induced by the mediator of its interactions with the Standard Model.
High-energy neutrinos constitute one of the very few direct accesses to energy scales above a few TeV. 
An indirect search for secluded DM signals has been performed with the ANTARES neutrino telescope using data from 2007 to 2015. Upper limits on the DM annihilation cross section for DM masses up to 6 PeV are presented and discussed.

\end{abstract}

\newpage

\tableofcontents

\section{Introduction}

Astrophysical and cosmological observations point to the existence of non-luminous matter beyond that contained in the Standard Model (SM) of particle physics.
Among the many proposed candidates for such dark matter (DM),
weakly interacting massive particles, with a mass at the electroweak scale, have been long looked for.
They annihilate to ordinary particles detectable far from their source, are scattered by ordinary matter, and can be produced at colliders.
No clear evidence for their existence has emerged so far from data. This situation is encouraging the exploration of new regions of the DM parameter space, and indeed recent years have seen a growing theoretical interest in DM candidates heavier than about 10~TeV.
This mass range is of even more interest in light of the empty-handed searches for physics beyond the Standard Model at the LHC, which push new physics models at scales larger than a few TeV, see e.g.~\cite{atlas:summary, cms:publications}.
In turn, these models may naturally host dark matter candidates with a mass in a similar range, as known since a long time, for example in supersymmetric theories~\cite{Dimopoulos:1996gy}.

Considerations of unitarity of DM annihilation processes imply the existence of a well-known upper limit, of about 100~TeV, on the DM mass~\cite{Griest:1989wd}, see e.g.~\cite{vonHarling:2014kha,Smirnov:2019ngs} for recent appraisals.
This limit holds if some conditions about the cosmological history of the universe and of DM are respected, and can for example be easily evaded if the universe was matter dominated between the freeze-out of dark matter interactions and Big Bang nucleosynthesis, see e.g.~\cite{Giudice:2000ex}.
Secluded DM models~\cite{Pospelov:2007mp} naturally constitute a very economical framework that realises the needed early-matter domination~\cite{Berlin:2016vnh,Berlin:2016gtr,Cirelli:2016rnw,FilSpectra}. 
Here the dark matter particle interacts sizeably with a mediator, in turn feebly interacting with SM particles. 
In these scenarios, the unitarity bound on the mass of thermal dark matter is avoided thanks to the late time entropy injection from decays of the mediators, which are responsible for the early-matter domination.
Dark matter masses of 100~TeV and above are therefore allowed.
Such models provide large signals in the so-called indirect detection searches (because controlled by the dark matter-mediator interaction) with almost no signal in direct detection and collider experiments (because controlled by the small mediator-SM coupling).
From a technical point of view, a reliable phenomenological computation of the spectra of SM particles arising from DM annihilations in secluded model is possible, providing an excellent motivation
for indirect searches~\cite{FilSpectra}.
Indeed, the relevant energy scale is not the heavy dark matter mass (that would demand a resummation of electroweak radiation for\footnote{  Through all the text, units are chosen such that $c=\hbar=1$.}
$m_\text{DM} > O(10)$~TeV, see~\cite{Bauer:2020jay} for a recent study that addressed this challenge), but rather the sub-TeV mediator mass, where the first order treatment of electroweak corrections~\cite{Ciafaloni:2010ti} implemented in the tool PPPC4DMID~\cite{Cirelli:2010xx} is well under control.
Therefore, despite the absence of prior bounds on the mass of the mediator $m_V$ from theory, a reliable computation of the indirect detection signals with~\cite{Cirelli:2010xx} is only possible when $m_V < O(10)$~TeV,
otherwise electroweak radiation should be resummed also to compute decays of the mediator.
This condition on the masses also implies that the interaction between DM particles is long-range, giving rise to phenomena like Sommerfeld enhancement~\cite{Sommerfeld:1931,Sakharov:1948yq} and bound state formation~\cite{Pospelov:2008jd,MarchRussell:2008tu,Shepherd:2009sa}, which significantly enhance the DM signal at present times with respect to the `standard' case of short-range interactions.

Neutrino telescopes have been used for indirect searches 
of DM (see for instance \cite{2021Univ....7..415Z} for a recent review).
The ANTARES detector has been used before to search
for DM accumulated in the Earth \cite{2017PDU....16...41A}, the Sun \cite{2016PhLB..759...69A} and the Galactic Centre \cite{2020PhLB..80535439A}. Moreover, there has been a specific search for secluded DM with ANTARES looking at the Sun \cite{2016JCAP...05..016A}, and in the public data from IceCube \cite{2017JCAP...04..010A, 2019JCAP...11..011N}. However, the Sun is not the best source to explore heavy DM  due to absorption of resulting particles in this dense medium, even high-energy neutrinos. Thus, it seems for this case more appropriate to look at the Galactic Centre and 
high-energy neutrinos constitute one of the very few direct accesses to energy scales above a few TeV. This places the ANTARES telescope in a privileged position to test this relatively unexplored mass range for dark matter, via the search for neutrinos possibly coming from dark matter annihilations or decays. This position is reinforced by the favourable geographical location of the telescope with respect to the position of the Galactic Centre, where most of the indirect signal from dark matter is expected to originate. 
It appears therefore very well motivated to exploit ANTARES data to test models of dark matter heavier than a few~TeV.

This paper is organised as follows. The production spectra for heavy secluded dark matter are detailed in Section~\ref{sec:spectra}. A description of the experimental setup is presented in Section~\ref{sec:detector} (as for the detector and data set used) and~\ref{sec:analysis} (as for the analysis method). The results of this work are exposed and discussed in Section~\ref{sec:results}, summarised and placed in further context in Section~\ref{sec:conclusions}.

\section{Neutrinos from Dark Matter annihilations}
\label{sec:spectra}

The neutrino signal at the ANTARES site arises from the annihilation of a pair of dark matter particles into two mediators.
They then decay into neutrinos and/or other SM particles, which in turn will produce neutrinos via showering and decays.
The mediator lifetime is required to be shorter than about 0.1 seconds to respect limits from Big Bang nucleosynthesis~\cite{Jedamzik:2006xz}.
With this constraint, the mediator decay process is instantaneous from the astrophysical point of view, and takes place entirely in the source of interest. The baryonic matter density in the Galactic Centre is not enough to cause distortions or absorption effects in outcoming neutrino spectra.
The formation of positronium-like bound states of DM can sizeably contribute to the signal of interest for ANTARES, via the decay of the bound state into two or more mediators~\cite{Pospelov:2008jd}. The dark matter annihilation cross section, for which limits will be presented here, is then to be intended as an effective cross section taking into account also the bound state contribution (see e.g.~\cite{Petraki:2016cnz,Cirelli:2016rnw} for more details).

\begin{figure}[h!]
    \centering
    \includegraphics[width=0.48\textwidth]{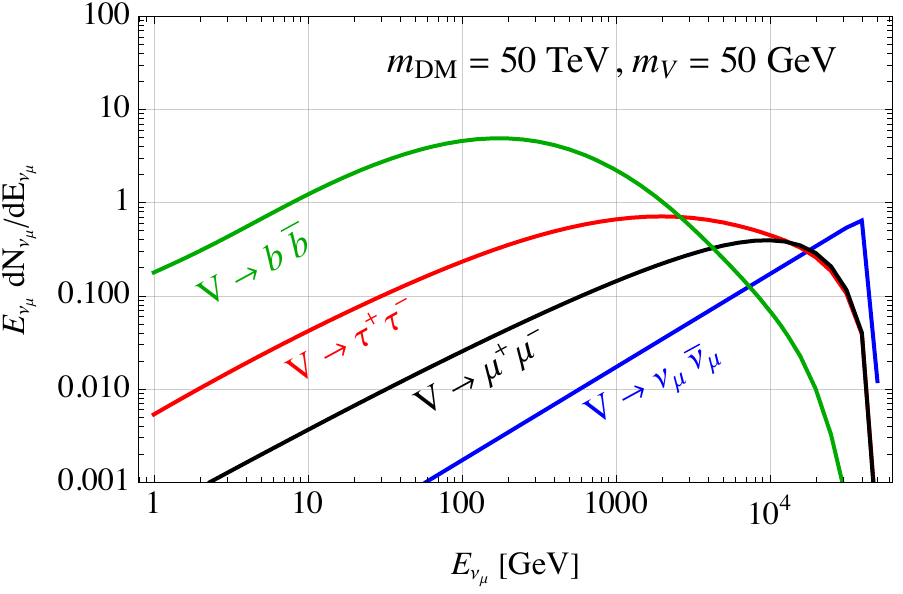}
    \;
    \includegraphics[width=0.48\textwidth]{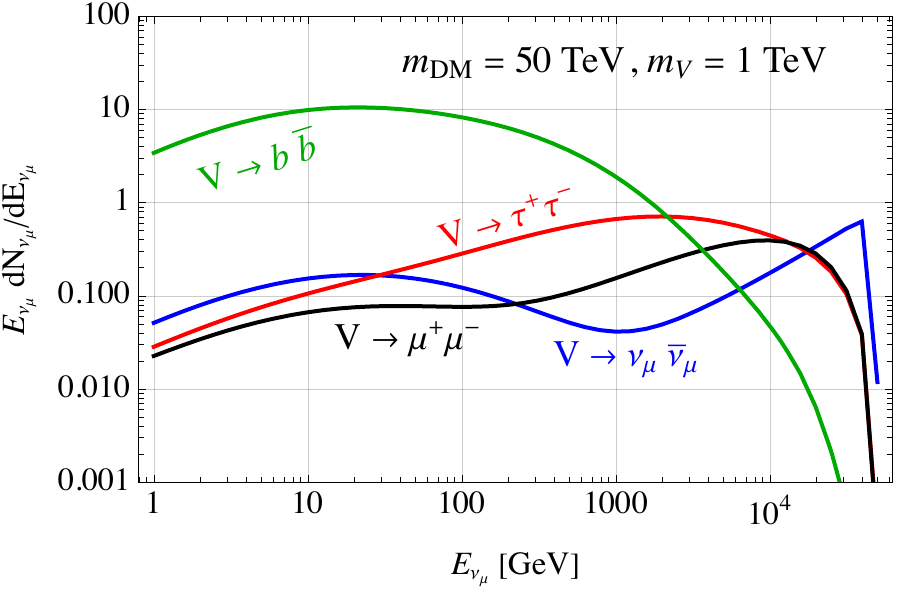}
    \caption{
    Energy distribution of the muon neutrinos plus antineutrinos at Earth location, per single annihilation into two mediators $V$ of a pair of DM particles each with mass of 50~TeV. 
The mediator decays to the SM pair indicated in the legend, then all (anti)neutrino flavours coming from that specific pair are included and contribute via long-distance oscillations to the muon (anti)neutrinos at Earth location. The mediator mass is 50 GeV in the left-hand plot and 1 TeV in the right-hand one.}
\label{fig:spectraFil}
\end{figure}

The energy spectra of the neutrinos per single dark matter annihilation are computed in two steps.
First, the energy spectra of neutrinos from the decay of a mediator at rest are obtained with the PPPC4DMID tool~\cite{Cirelli:2010xx}. Second, spectra are boosted to the centre of mass frame of the dark matter pair that annihilates (see~\cite{Elor:2015bho} for more details on this procedure).
Flavour oscillations then occur between the source and the detector site. In this analysis, the production of three neutrino flavours was considered in the Galactic Centre, and oscillated in the long-baseline approximation to obtain spectra at the Earth surface. Figure~\ref{fig:spectraFil} shows these spectra for two benchmark values of the mediator mass $m_V$, and for a DM mass of 50 TeV.
When $m_V = 50$~GeV, electroweak corrections to the spectra are not important.
Considering as an example the $V \to \nu_\mu \bar{\nu}_\mu$ channel, one can then understand the shape of its spectrum as follows:
in the mediator frame, the spectrum consists of a delta-function, the energy of each neutrino is half the mass of the mediator. When this delta is boosted to the frame of the DM pair, it gives rise to neutrinos spread over all energies and up to the DM mass, as visible in the left-hand panel of Figure~\ref{fig:spectraFil}.
Instead, when $m_V = 1000$~GeV electroweak corrections are important, and they are for example responsible for the ``bump'' visible at low energies in the $V \to \nu_\mu \bar\nu_\mu$ channel: the neutrinos from the decay of the mediator can radiate a $W$ or a $Z$ boson, which in turn will give rise to more neutrinos at smaller energies.
Analogous considerations apply to the other $V$ decay channels.
Apart from oscillation effects, the primary energy spectra above coincide with the spectra at the ANTARES location, as neutrinos of these energies propagate undisturbed in the Galaxy.
In this analysis, the following decay channels of mediators $V$ into Standard Model particles have been considered:
\begin{equation}
    V \rightarrow \mu^+\mu^-, \; \tau^+\tau^-, \; b\bar{b}, \; \nu_\mu \bar{\nu}_\mu\,.
\end{equation}
Each of these channels is treated independently with a branching ratio of 100\%.

\section{Detector and data set}
\label{sec:detector}
The ANTARES neutrino detector is situated underwater in the Mediterranean Sea 40 km offshore from Toulon (France). It is composed of 12 lines instrumented with photomultiplier tubes for the detection of Cherenkov light \cite{ANTARES}. 
ANTARES	records Cherenkov light induced by charged particles originated in the interaction of a neutrino inside the detector or in the volume around it. Based on these recorded signals, the neutrino energy and arrival direction are reconstructed and constitute the main information of processed data.
In the text that follows the term {\em neutrinos} stands for $\nu$ and ${\bar{\nu}}$, as the	events	generated	by	their	 interactions	are	seen	indistinguishably in neutrino telescopes.
Muons produced in cosmic ray interactions in the atmosphere form a very large background which is suppressed in analyses by considering only events with arrival directions crossing the Earth.
The Galactic Centre, located at a declination of $-29.01^\circ$S, is visible from the detector latitude about 70$\%$ of the time \cite{Visibility}.
In this analysis, 9 years of muon tracks, mostly induced by upward-going $\nu_\mu$ charged current (CC) interactions and collected between May 2007 and December 2015, were searched. 
This sample is composed of 7637 reconstructed tracks recorded over 2101.6 days of effective livetime.
Tracks are reconstructed with a good angular resolution of the order of 1$^\circ$ at the energies relevant for this search \cite{AntaresAF};
this data set coincides with the one analysed in previous works \cite{ANT9YEARS}.

Tracks are reconstructed from the calibrated positions~\cite{2012JInst...7T8002A} and calibrated hit times~\cite{ANTARES:2010iah} of photomultiplier hits recorded in coincidence with the event.
A quality parameter $\Lambda$ is associated to each reconstructed track, based on a maximum likelihood obtained for the reconstruction fit \cite{ANTARES:2012xie}. 
In its geometrical layout, the ANTARES detector is designed for the detection of astrophysical neutrino fluxes, which ensures a good coverage of the energy range necessary for neutrinos from heavy dark matter annihilation.
The	amount of Cherenkov	photons	induced	in the paths of the propagating charged particles is	proportional	to	the	amount of deposited energy	and, consequently, the	number	of	hit	optical	modules,  $N_{\mbox{\tiny{HIT}}}$,	 is	a proxy	of	the neutrino energy	$E_\nu$.
A set of simulated data has been produced in correspondence with the environmental and trigger conditions of each ANTARES data run \cite{2021JCAP...01..064A}.
To reproduce the expected signal from secluded dark matter, the simulated event energy is 
weighted with a factor obtained according to the energy distributions of each annihilation channel computed following~\cite{FilSpectra} and shown in Figure \ref{fig:spectraFil}.

\section{Analysis method}
\label{sec:analysis}
The signature of secluded dark matter annihilation would be, as other dark matter signals, very difficult to distinctively identify. 
In this analysis, a stricter event selection has been applied with respect to previous searches for weakly interacting massive particles~\cite{2020PhLB..80535439A}, to setup a more assertive test of the non-standard scenario, as detailed in Section \ref{dataselection}. 
With the preliminary event selection described in this section, the sample is cleaned off the majority of atmospheric muons mis-reconstructed as upgoing that failed to be removed by standard analysis cuts.
The remaining atmospheric neutrinos plus a possible component from dark matter annihilations compose our `pre-selected' data sample. 
An unbinned maximum likelihood method is applied to this pre-selected sample to search for signals of secluded dark matter over the underlying background of atmospheric neutrinos. The discrimination between atmospheric neutrinos and neutrinos from dark matter annihilation is based on a space and morphology information on the location of the source, and on a spectral information based on the knowledge of the energy distribution of each DM annihilation channel. This method has been used in previous analyses such as \cite{ANT9YEARS, 2020PhLB..80535439A}.

\subsection{Data Selection}\label{dataselection}
\begin{figure}
\includegraphics[width=0.49\textwidth]{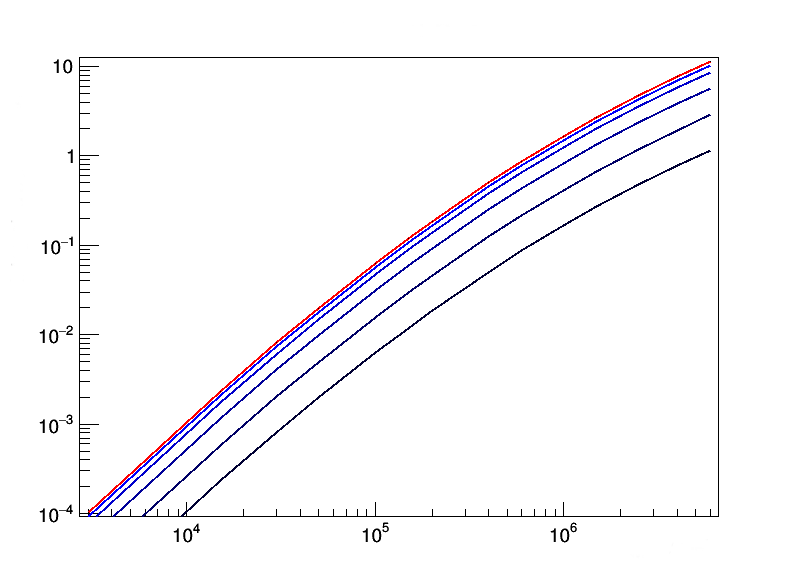}  
\put(-200, 20){\rotatebox{90}{\scriptsize{Integrated acceptance [m$^2$]}}}
\put(-140,110){\scriptsize{$b$ channel}}           
\includegraphics[width=0.49\textwidth]{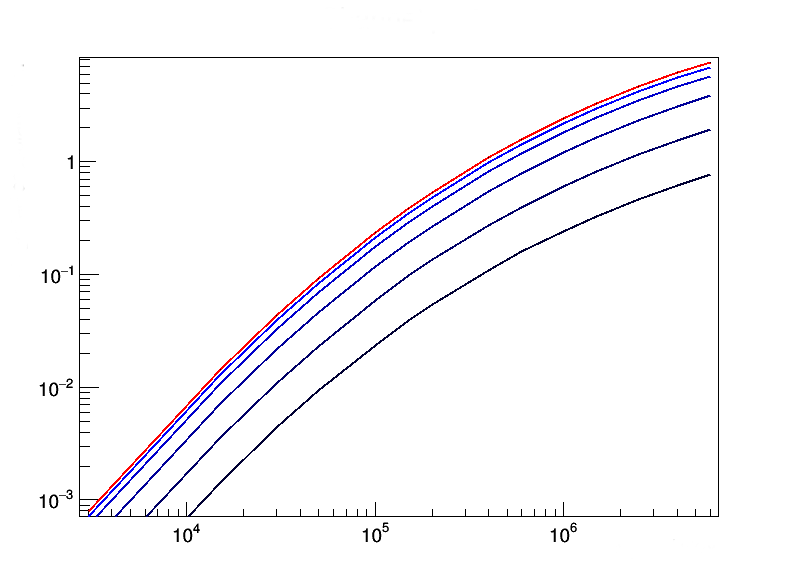}  
\put(-140,110){\scriptsize{$\mu$ channel}}           
\put(-200, 20){\rotatebox{90}{\scriptsize{Integrated acceptance [m$^2$]}}}\\
\includegraphics[width=0.49\textwidth]{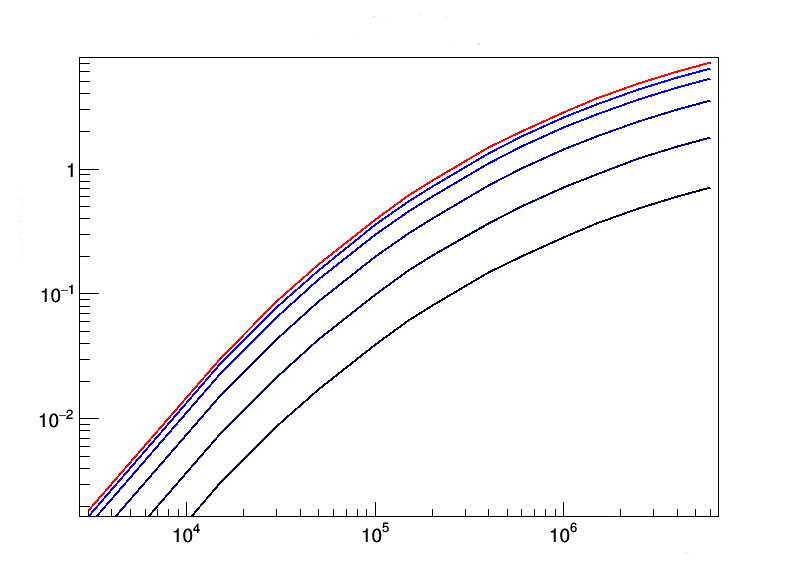}
\put(-140,110){\scriptsize{$\nu_\mu$ channel}}           
\includegraphics[width=0.49\textwidth]{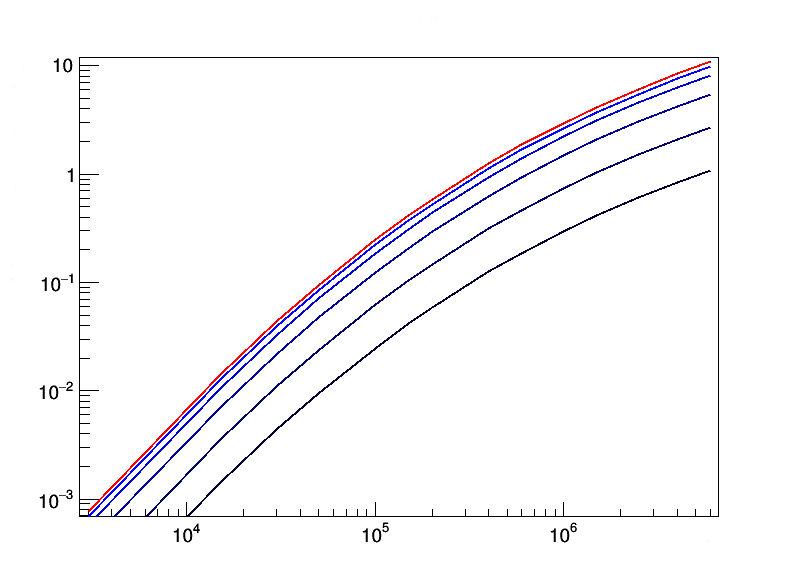}
\put(-70,-5){\scriptsize{$m_{\textrm{DM}}$ [GeV]}}           
\put(-70,150){\scriptsize{$m_{\textrm{DM}}$ [GeV]}}           
\put(-270,-5){\scriptsize{$m_{\textrm{DM}}$ [GeV]}}           
\put(-270,150){\scriptsize{$m_{\textrm{DM}}$ [GeV]}}
\put(-140,110){\scriptsize{$\tau$ channel}}
\put(-200, 20){\rotatebox{90}{\scriptsize{Integrated acceptance [m$^2$]}}}
\put(-400, 20){\rotatebox{90}{\scriptsize{Integrated acceptance [m$^2$]}}}
    \caption{Integrated acceptances, as defined in Equation~\ref{eqacc}, computed for four annihilation spectra (indicated over each panel). Acceptances computed without any $N_\mathrm{HIT}$ cut are shown with a red line; different cut values in $N_\mathrm{HIT}$ as reported in Table \ref{NHITtable}, leading to a reduction of acceptance from 90\% to 10\%}, are indicated with blue shades from brighter to darker respectively.
    \label{FigAcceptance}
\end{figure}
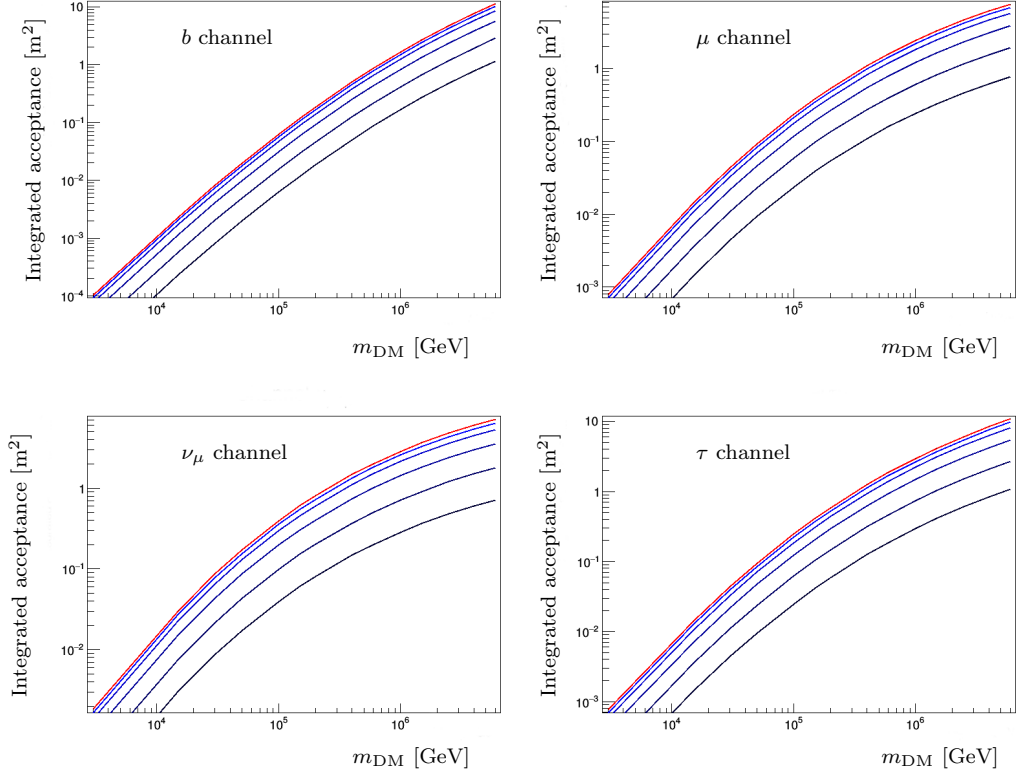

A set of relaxed starting cuts is initially applied to the data sample in order to reduce a large fraction of background from atmospheric muons, and perform consistency checks between data and Monte Carlo simulation. Similarly to other DM analyses by the ANTARES Collaboration~\cite{2020PhLB..80535439A,ANTARES:2012xie,ANT9YEARS}, these cuts regard quality indicators of the reconstructed events: the likelihood $\Lambda$ for the linear fit interpolating the hit pattern, and the angular uncertainty $\beta$ estimating the angular error on the track arrival direction. The condition for the reconstructed event to be coming from across the Earth is required with a cut on $\theta$, zenith angle of the  reconstructed track (with respect to an axis pointing up to the vertical).
Initially, events fulfilling $ \Lambda > -5.6$, $\beta < 1^\circ$ and $\theta< 90^\circ$ are selected. With these starting cuts, the suppression of atmospheric muons, electronic noise and poorly reconstructed tracks is ensured. A stricter set of variable cuts is then applied as summarised in what follows. 
The data selection was optimised by maximizing the sensitivity on the signal by varying $\Lambda$ (from $-5.4$ to $-4.8$ in steps of 0.2).
A cut on the neutrino energy is introduced, motivated by the shape of the energy distribution of annihilating secluded dark matter (see Figure \ref{fig:spectraFil}), which favours events in the high-energy end of the spectrum. On the contrary, a power law describes the atmospheric neutrino spectrum. 
In this analysis the number of recorded light hits $N_{\mathrm{HIT}}$ is used as a proxy for the reconstructed neutrino energy. In addition to the cuts on zenith angle, $\Lambda$ and $\beta$, $N_{\mathrm{HIT}}$ is varied up to 200 in steps of 1.
The cut value leading to the strongest sensitivity in velocity averaged cross section for DM annihilation
$\langle \sigma v \rangle$ is chosen and applied to the unblinded data set.
While the best value for $\Lambda$ remains fixed at $-5.2$, consistently with previous similar analyses \cite{ANT9YEARS}, the cut value on $N_{\mathrm{HIT}}$ varies according to the dark matter mass $m_{\mathrm{DM}}$, annihilation channel, and mediator mass $m_V$; the corresponding values are reported in Table~\ref{NHITtable}.
The flux of signal events is obtained dividing the number of signal events (or limit) by the integrated acceptance, defined as the integral of the effective area $A_{\mathrm{EFF}}$ weighted by the dark matter annihilation spectrum
\begin{equation}\label{eqacc}
    \mathcal{A}(m_\mathrm{DM}) = \int_{0}^{m_\mathrm{DM}}  A_\mathrm{EFF}^{\nu}(E_\nu) \frac{dN_\nu (E_\nu)}{dE_\nu} dE_\nu + A_\mathrm{EFF}^{\bar\nu}(E_\nu) \frac{dN_{\bar\nu} (E_{\bar\nu})}{dE_{\bar\nu}} dE_{\bar\nu}.
\end{equation}
Tightening any cut improves the purity of the signal sample but reduces the acceptance. Initially, values for the $N_{\mathrm{HIT}}$ cuts have been chosen such to reduce the acceptance to 90\%, 75\%, 50\%, 25\%, 10\% with respect to the uncut value (where uncut means including all other cuts). However, the values 25\% and 10\% result in a suppression of too many events for performing the likelihood analysis. 
In these cases, it was impossible to successfully scan the skymap looking for a signal cluster, and therefore these cut values were not further considered in the analysis. 
The corresponding acceptances are shown in Figure \ref{FigAcceptance} for the four annihilation modes $V\rightarrow b \bar{b} $, $V\rightarrow \mu^+\mu^-$, $V\rightarrow \nu_\mu \bar\nu_\mu$ and $V\rightarrow \tau^+\tau^-$.

\begin{table}[h!]
\begin{scriptsize}
$$
{\begin{array}{cc|ccccccccc|cccccc|}
&&&&&&&& m_{\mathrm{DM}} &&&&&&\\
&&&&&&  \mathrm{TeV} &&&&&&& \mathrm{  PeV} &\\
&{\mathrm{channel}}&  3  & 15  & 30  & 50  & 100   & 150  & 200   &  400  &  600  &  1  & 1.5  & 2.5  & 4  & 6\\
\hline
\hline
&\mu &31&33&35&36&38&39&40&77&82&87&92&97&102&106\\
m_V = &\tau &31&33&34&35&37&38&39&74&77&82&86&91&96&99\\
\mathrm{50\,GeV}&b & 29&31&32&32&33&34&35&36&37&38&39&40&75&78\\
&\nu_\mu&31&34&36&38&40&75&78&86&91&97&102&107&111&113\\
\hline
&\mu &31&33&35&36&38&39&40&77&82&87&92&97&102&106\\
m_V = &\tau &31&33&34&35&37&38&39&74&77&82&86&91&96&99\\
\mathrm{250\,GeV}&b &29&31&32&32&33&34&35&36&37&38&39&71&75&78\\
&\nu_\mu &31&34&36&38&70&75&78&86&91&97&102&107&111&113\\
\hline
&\mu &31&33&35&36&38&39&52&77&82&87&92&97&102&106\\
m_V = &\tau &31&33&34&35&37&38&39&74&77&82&86&91&96&99\\
\mathrm{1\,TeV}&b &29&31&32&32&33&34&35&36&37&38&39&71&75&78\\
&\nu_\mu &31&34&36&38&40&75&78&86&91&97&102&107&111&113\\
\hline\\
 \end{array}}
 $$
 \end{scriptsize}
\caption{Cut values on the number of hits $N_{\mathrm{HIT}}$, optimised for best sensitivity for each dark matter mass $m_{\mathrm{DM}}$, annihilation channel, mediator mass $m_V$.}
\label{NHITtable}
\end{table}

The search for a signal of secluded dark matter is optimised in a $blind$ way, according to which the  events are shuffled in right ascension to ensure the unbiased optimisation of selection cuts. When blinding, a random number $\alpha_{\mathrm{blind}} \in [-180^\circ, 180^\circ]$ is assigned as right ascension of the reconstructed arrival direction. After establishing the best sensitivities, the original right ascension coordinate is set back. The blinding procedure does not alter the expected sky distribution of atmospheric neutrinos, in which the declination coordinate is maintained.

\subsection{Signal Identification}
Based on the information about the expected signal, an unbinned maximum likelihood algorithm has been used as a search method. Unbinned likelihood is a fitting method based on the prior knowledge of probability distribution functions (PDFs) of signal and background discriminating variables. The method used here does not differ from the one already applied in other ANTARES analyses such as \cite{ANT9YEARS, 2020PhLB..80535439A}, and comprehends the following steps:
\begin{enumerate}
    \item Computation of PDFs for the signal, based on the spectra described in~\cite{FilSpectra} and DM halo morphology described with Navarro–Frenk–White (NFW) profile of spatial mass distribution of DM profile~\cite{NFW} $\rho_{DM}(r) = \rho_s\; r_s/(r (1+r^2/r_s^2))$, with $\rho_s = 1.40 \cdot 10^7 M_\odot$/kpc$^3$ and $r_s = 16.1$ kpc \cite{2020PhLB..80535439A}. 
    Computation of PDF for the distribution of atmospheric background events is obtained from blind data. 
    \item Generation of $10^4$ pseudo-skymaps for each DM parameter choice ($m_\mathrm{DM}$, $m_V$, annihilation channel). A number of signal events scanned in steps of 1 between 0 and 50 is injected in addition to the total number of events taken from the background sample. Each choice of a number of signal events makes a population of pseudo-skymaps.
    \item Maximisation of likelihood yielding a test statistic (TS) distribution  for each population of pseudo-skymaps. A convolution with a Poisson function is performed to include fluctuations expected in the distribution of signal events.
    \item Computation of 90\% confidence level (CL) median upper limit in number of detectable events that will be referred to as sensitivity according to the Neyman method~\cite{Neyman:1938}. The flux sensitivity is obtained dividing the sensitivity on number of events by the integrated acceptance.
    \item {\em{Unblinding}}: determine the likelihood of the real data distribution and, if no evidence of excess,  computation of limits at 90\% CL on flux and velocity averaged annihilation cross section $\langle \sigma v\rangle$. These results are presented in Section~\ref{sec:results}.
\end{enumerate}
In order to quantify the signal component, a TS is defined as the ratio between the maximum likelihood and the likelihood of the pure background sample. Sensitivities at 90\% CL are obtained comparing the TS distribution for different numbers of injected signal events with the median of pure background distribution, and selecting the population which is confused with background less than 10\% of the times. 
The number of events $n_s^*$  reconstructed with maximum likelihood in each set of pseudo-skymaps is subject to fluctuations following a Poisson distribution. To include fluctuations, a transformation through a Poisson function, $\mathcal{P}$, is performed, returning the TS distribution $P(\mathrm{TS})$ as a function of the Poissonian mean $\mu$:
\begin{equation}
P\left(\mathrm{TS} (\mu)\right) = \sum_{n^*_s=1}^N P\left(\mathrm{TS}(n_s^*)\right)  \,\mathcal{P} (n_s^*, \mu).
\end{equation}
As in similar analyses \cite{ANT9YEARS}, 
to take into account systematics on the expected number of $\nu_\mu$~CC reconstructed events, a smearing of the test statistic with a 15\% width Gaussian is performed \cite{AntaresAF}.

\section{Results and discussion}
\label{sec:results}
\begin{figure}
\centering
\includegraphics[width=0.85\columnwidth]{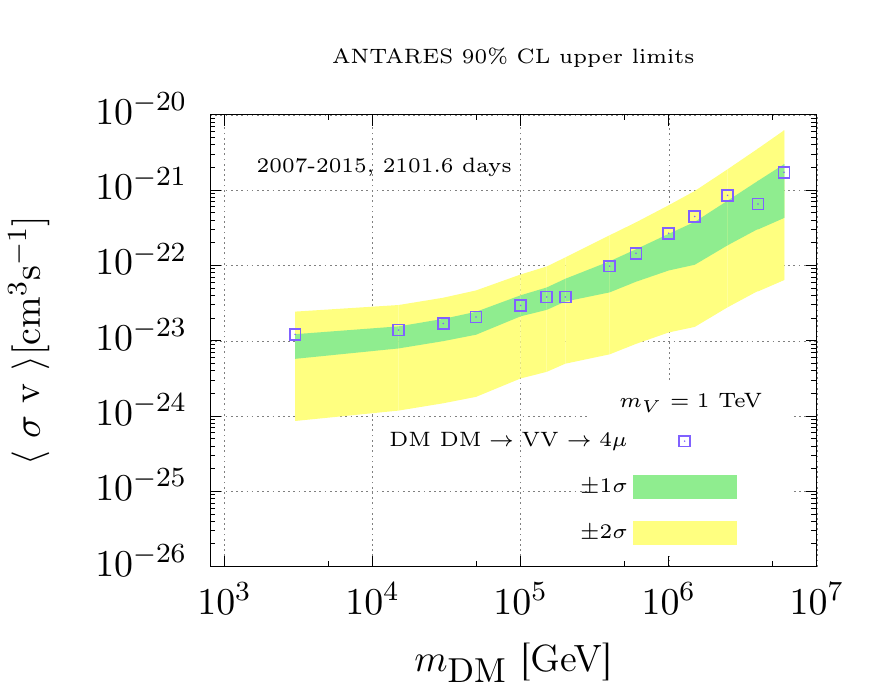}
\includegraphics[width=0.85\columnwidth]{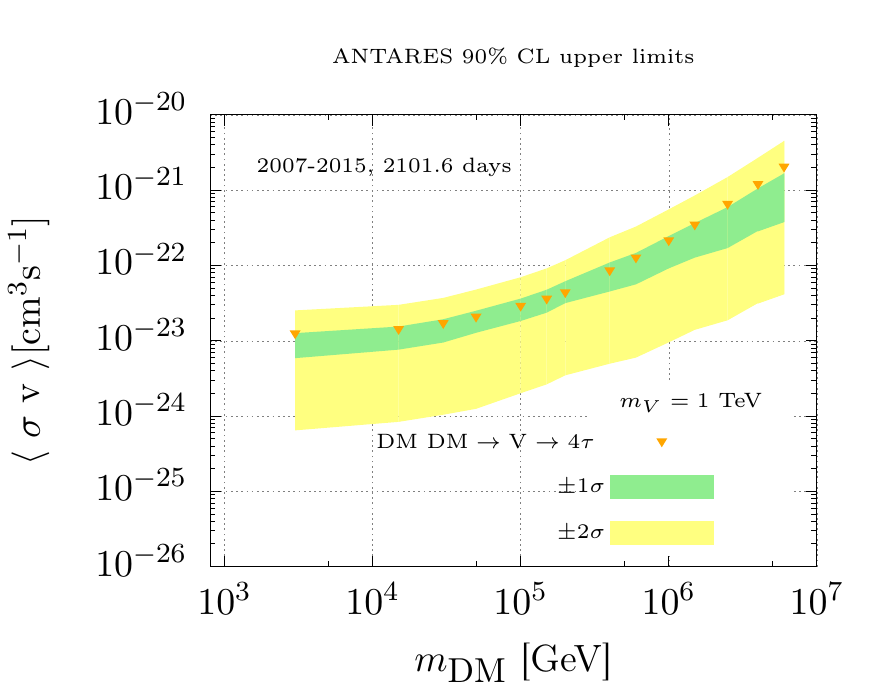}
\caption{\label{bands}Upper limits at 90\% CL  on the thermally averaged DM pair annihilation cross section $\langle \sigma v \rangle$ for a mediator mass $m_V$= 1 TeV, with 1$\sigma$ and 2$\sigma$ containment bands, for $4\mu$ (top panel, limits as blue boxes) and $4\tau$ (bottom panel, limits as orange triangles) final states.}
\end{figure}

\begin{figure}
\centering
\includegraphics[width=0.85\columnwidth]{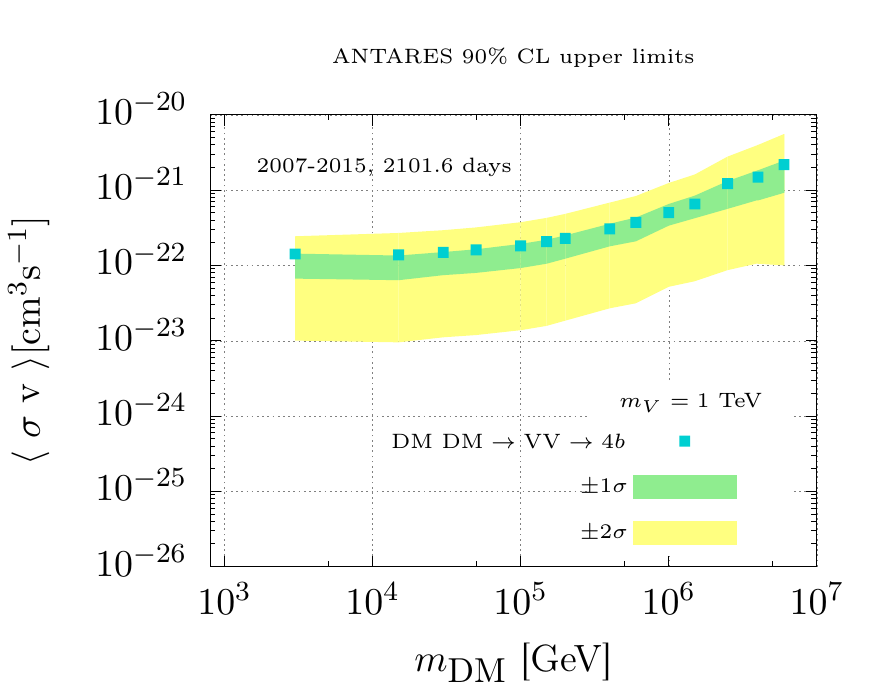}
\includegraphics[width=0.85\columnwidth]{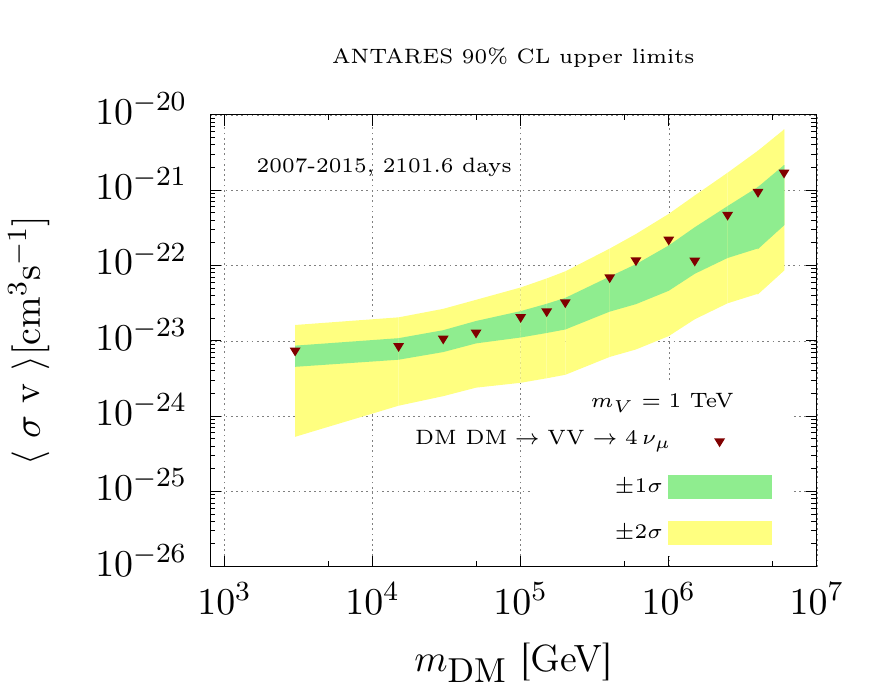}
\caption{\label{bands2} Upper limits at 90\% CL  on the thermally averaged DM pair annihilation cross section $\langle \sigma v \rangle$ for a mediator mass $m_V$= 1 TeV, with 1$\sigma$ and 2$\sigma$ containment bands, for $4b$ (top panel, limits as cyan boxes) and $4\nu$ (bottom panel, limits as brown triangles) final states.}
\end{figure}

This search for heavy secluded dark matter is performed as a function of three free parameters: the dark matter candidate mass $m_\mathrm{DM}$, the mediator mass $m_V$ (with general condition $m_V \ll m_\mathrm{DM}$) and the annihilation channel. As mentioned before, the galactic DM halo profile
 has been fixed to the NFW parameterisation~\cite{NFW}. As explained in the previous section, a set of optimal cuts, identified independently for each parameter choice (14 dark matter masses, 3 mediator masses and four annihilation channels) is applied to the data in $14 \times 3 \times 4$ unblindings. 
Data is found to be consistent with the background-only hypothesis.
Upper limits at 90\% CL on the thermally averaged cross section for self-conjugate DM pair annihilation are computed for a light ($m_V = 50$ GeV), a medium ($m_V = 250$ GeV) and a heavy ($m_V = 1$ TeV) mediator.
The most stringent limits are obtained in the direct channel $\mathrm{DM}\,\mathrm{DM}\rightarrow2\,V\rightarrow 4\nu$, which is due to the spectral shape of the 4$\nu$ annihilation mode, which among those considered yields the largest fraction of neutrinos at high energies. 
Each annihilation mode is independently considered with a branching ratio of 100\%.
Figures~\ref{bands} and~\ref{bands2} display the results of the upper limits for each channel separately, alongside with the sensitivities and corresponding $1\sigma$ and $2\sigma$ containment bands shaded in green and yellow respectively.
1000 sets of simulated data are generated and used
to determine the sensitivity as the mean expected exclusion, and the bands as the 680 and 950 closest lines to the sensitivity.
This procedure allows one to visualise possible statistical fluctuations of the background: the fact that our limits stay within the bands means that data are compatible with the background hypothesis at better than $2\sigma$. 
Limits have been raised to be equal to sensitivities in case of underfluctuations, analogously to similar ANTARES analyses \cite{ANT9YEARS}.

\begin{figure}
\centering
\includegraphics[width=0.9\columnwidth]{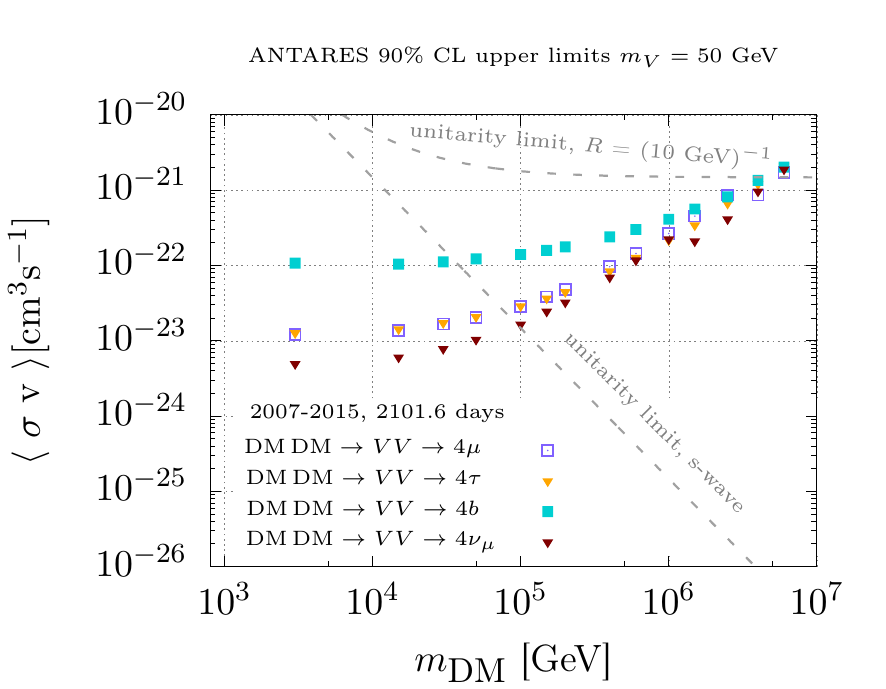}\,
\caption{\label{limits} Upper limits at 90\% CL on the thermally averaged cross section for DM pair annihilation $\langle \sigma v \rangle$, from the analysis of 9 years of ANTARES data, for mediator masses $m_V$= 50 GeV, and for the mediator decay channels 4$\mu$, 4$\tau$, 4$b$,  4$\nu_\mu$. The dashed lines denote the unitarity limit on the DM annihilation cross section in two limiting cases, one where only the $s$-wave dominates the scattering, and one where DM is a composite object with size $R \simeq (10~\text{GeV})^{-1}$. In each of these cases, the parameter space above the related line is theoretically inaccessible, see text for more details.} 
\end{figure}

\begin{figure}
\centering
\includegraphics[width=0.9\columnwidth]{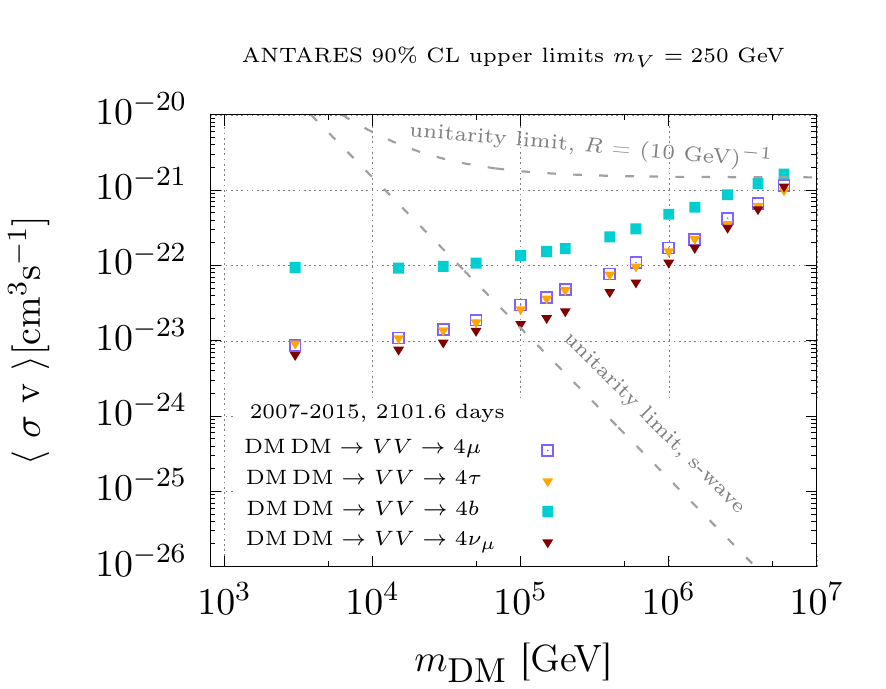}\\
\includegraphics[width=0.9\columnwidth]{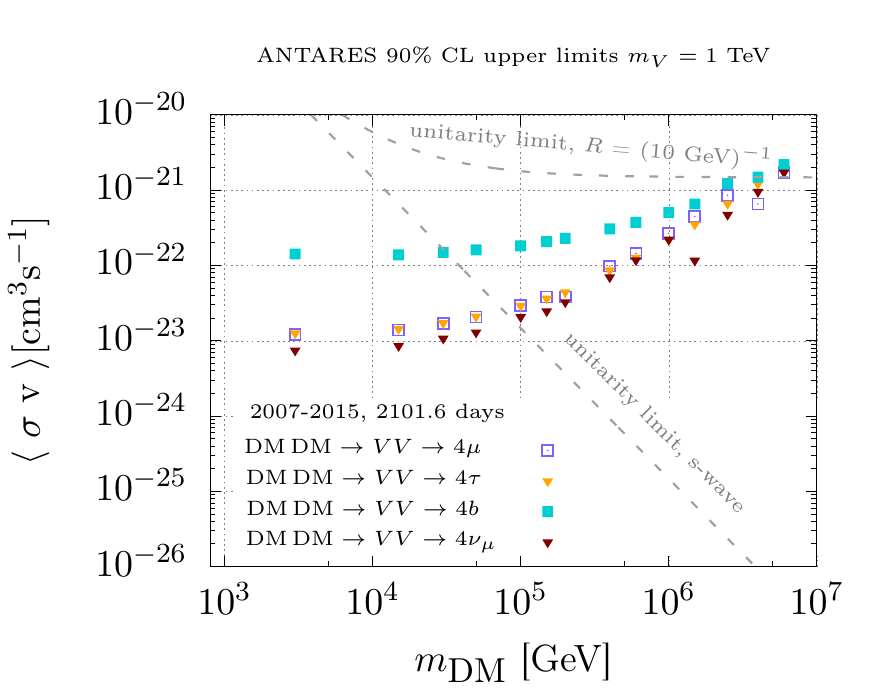}
\caption{\label{limits2} Same as Figure \ref{limits} for mediator masses $m_V$= 250 GeV (top panel) and 1~TeV (bottom panel).} 
\end{figure}

To understand which DM models are tested,
Figures~\ref{limits} and~\ref{limits2} display also the upper limits together with the two lines indicating the unitarity limit on the annihilation cross section. They assume, respectively, that
annihilation is dominated by s-wave processes~\cite{Griest:1989wd,Smirnov:2019ngs}
\begin{equation}
  \sigma v < \frac{4 \pi}{v}\frac{1}{m_\text{DM}^2}  ,
  \label{eq:uni_s}
\end{equation}
or that DM is a composite state with size $R \simeq (10~\text{GeV})^{-1}$
\begin{equation}
   \sigma v < \frac{4 \pi}{v}\frac{1}{m_\text{DM}^2} \Big(1+m_\text{DM} v R\Big)^2\,.
   \label{eq:uni_composite}
\end{equation}
The extra term in Eq.~(\ref{eq:uni_composite})  with respect to Eq.~(\ref{eq:uni_s}) is the result of the sum over all partial waves $\sum_{j=0}^{j_\text{max}}(2j+1)$, where $j_\text{max} = m_\text{DM} v R$, see e.g.~\cite{Griest:1989wd,Smirnov:2019ngs} for more details.
These two lines should be regarded as rough indications of which models are tested by the searches presented in this paper. One for example learns that the DM models, for which DM masses heavier than 100 TeV can be tested, are those where more than a single partial wave contributes significantly to the annihilation cross section. Composite DM is a limiting case where a large number of partial waves contributes, see e.g.~\cite{Mitridate:2017oky,Contino:2018crt,Geller:2018biy}.
Other DM models that evade the unitarity limit feature an indirect detection phenomenology analogous to the one of secluded models (e.g., supercooled composite DM~\cite{Baldes:2020kam,Baldes:2021aph}), so they are also constrained by the searches presented here. The interest of the limits presented in this paper goes therefore beyond the cosmological histories with early matter domination, sketched in the introduction.

\section{Conclusions}
\label{sec:conclusions}
A search in ANTARES data from 2007 to 2015 for a signal, coming from the Galactic Centre, due to the annihilation of secluded dark matter particles was presented. Data were found to be consistent with the background-only hypothesis, so that limits on the velocity averaged cross sections for annihilation were placed for DM candidate masses between 3 TeV and 6 PeV. These limits have been compared with theoretical expectations for the maximal possible annihilation signals in different models.

Previous DM searches with ANTARES have used the information on the energy of each event (i.e. the number of hits) as an input variable for the likelihood, and then computed limits integrating between the minimal ANTARES sensitivity and the largest energy allowed by the signal model~\cite{2020PhLB..80535439A}.
This search instead constitutes the first case where the information on the energy is used to preselect events, namely the lowest energy has been varied with the parameters of the signal model tested, to optimise the sensitivity of ANTARES in testing it.
To the best of our knowledge, this also constitutes the first time that any telescope tested annihilation signals from DM with masses up to the PeV range.

\section*{Acknowledgements}
The authors acknowledge the financial support of the funding agencies:
Centre National de la Recherche Scientifique (CNRS), Commissariat \`a
l'\'ener\-gie atomique et aux \'energies alternatives (CEA),
Commission Europ\'eenne (FEDER fund and Marie Curie Program),
Institut Universitaire de France (IUF), LabEx UnivEarthS (ANR-10-LABX-0023 and ANR-18-IDEX-0001),
R\'egion \^Ile-de-France (DIM-ACAV), R\'egion
Alsace (contrat CPER), R\'egion Provence-Alpes-C\^ote d'Azur,
D\'e\-par\-tement du Var and Ville de La
Seyne-sur-Mer, France;
Bundesministerium f\"ur Bildung und Forschung
(BMBF), Germany; 
Istituto Nazionale di Fisica Nucleare (INFN), Italy;
Nederlandse organisatie voor Wetenschappelijk Onderzoek (NWO), the Netherlands;
Executive Unit for Financing Higher Education, Research, Development and Innovation (UEFISCDI), Romania;
Ministerio de Ciencia, Innovaci\'{o}n, Investigaci\'{o}n y
Universidades (MCIU): Programa Estatal de Generaci\'{o}n de
Conocimiento (refs. PGC2018-096663-B-C41, -A-C42, -B-C43, -B-C44)
(MCIU/FEDER), Generalitat Valenciana: Prometeo (PROMETEO/2020/019),
Grisol\'{i}a (refs. GRISOLIA/2018/119, /2021/192) and GenT
(refs. CIDEGENT/2018/034, /2019/043, /2020/049, /2021/023) programs, Junta de
Andaluc\'{i}a (ref. A-FQM-053-UGR18), La Caixa Foundation (ref. LCF/BQ/IN17/
11620019), EU: MSC program (ref. 101025085), Spain;
Ministry of Higher Education, Scientific Research and Innovation, Morocco, and the Arab Fund for Economic and Social Development, Kuwait.
We also acknowledge the technical support of Ifremer, AIM and Foselev Marine
for the sea operation and the CC-IN2P3 for the computing facilities.
F.~Sala acknowledges funding support from the Initiative Physique des Infinis (IPI), a research training program of the Idex SUPER at Sorbonne Universit\'e.

\vskip1ex

\bibliographystyle{JHEP}
\bibliography{refs}

\end{document}